\documentclass[
reprint,
superscriptaddress,
longbibliography,
showkeys,
 amsmath,amssymb,
 aps,
pre, 
]{revtex4-2}

\usepackage{graphicx}
\usepackage[caption=false]{subfig} 

\usepackage{float}
     
\usepackage{dcolumn}
\usepackage{bm}

\newcommand{\ee}{\end{equation}} 
\newcommand{\be}{\begin{equation}}

\usepackage[mathscr]{euscript}  
\usepackage[dvipsnames]{xcolor}

\usepackage{tcolorbox}
\usepackage{comment}
\usepackage{float}

\makeatletter
\newsavebox{\@brx}
\newcommand{\llangle}[1][]{\savebox{\@brx}{\(\m@th{#1\langle}\)}%
  \mathopen{\copy\@brx\kern-0.5\wd\@brx\usebox{\@brx}}}
\newcommand{\rrangle}[1][]{\savebox{\@brx}{\(\m@th{#1\rangle}\)}%
  \mathclose{\copy\@brx\kern-0.5\wd\@brx\usebox{\@brx}}}
\makeatother

\begin{document} 

\preprint{APS/123-QED}

\title{Self-propulsion protocols for swift non-equilibrium state transitions\\ and enhanced cooling in active systems}

\author{Kristian St\o{}levik Olsen}
\thanks{\textit{Correspondence: kristian.olsen@hhu.de}}
\affiliation{Institute for Theoretical Physics II - Soft Matter, Heinrich-Heine University Düsseldorf, D-40225 Düsseldorf, Germany}

\author{Hartmut L\"{o}wen}
\affiliation{Institute for Theoretical Physics II - Soft Matter, Heinrich-Heine University Düsseldorf, D-40225 Düsseldorf, Germany}

\begin{abstract}
A control framework is proposed for inducing non-equilibrium state transitions in confined active matter, where the statistics of self‑propulsion serve as the only control parameter.  Positivity of the noise amplitudes and fundamental bounds on position–propulsion correlations define the admissible control space and impose speed‑limits on transitions between non‑equilibrium states. We show that non‑stationary initial states facilitate additional speed‑ups, corresponding to pre-loading the state with negative correlations. This enables {active} cooling protocols that outperform their passive counterparts.
\end{abstract}

\maketitle

\section{Introduction}

 Active matter systems exhibit diverse dynamical behaviors driven by continuous energy consumption at the particle level \cite{ramaswamy2010mechanics,marchetti2013hydrodynamics}. Unlike passive systems where dynamics is determined by thermal fluctuations, active systems are governed by internal driving mechanisms that generate transport properties and non‑equilibrium states without equilibrium counterparts. Understanding how time‑dependent driving protocols create and connect such states is therefore central to non‑equilibrium control.

\begin{figure}
    \centering
    \includegraphics[width= \columnwidth]{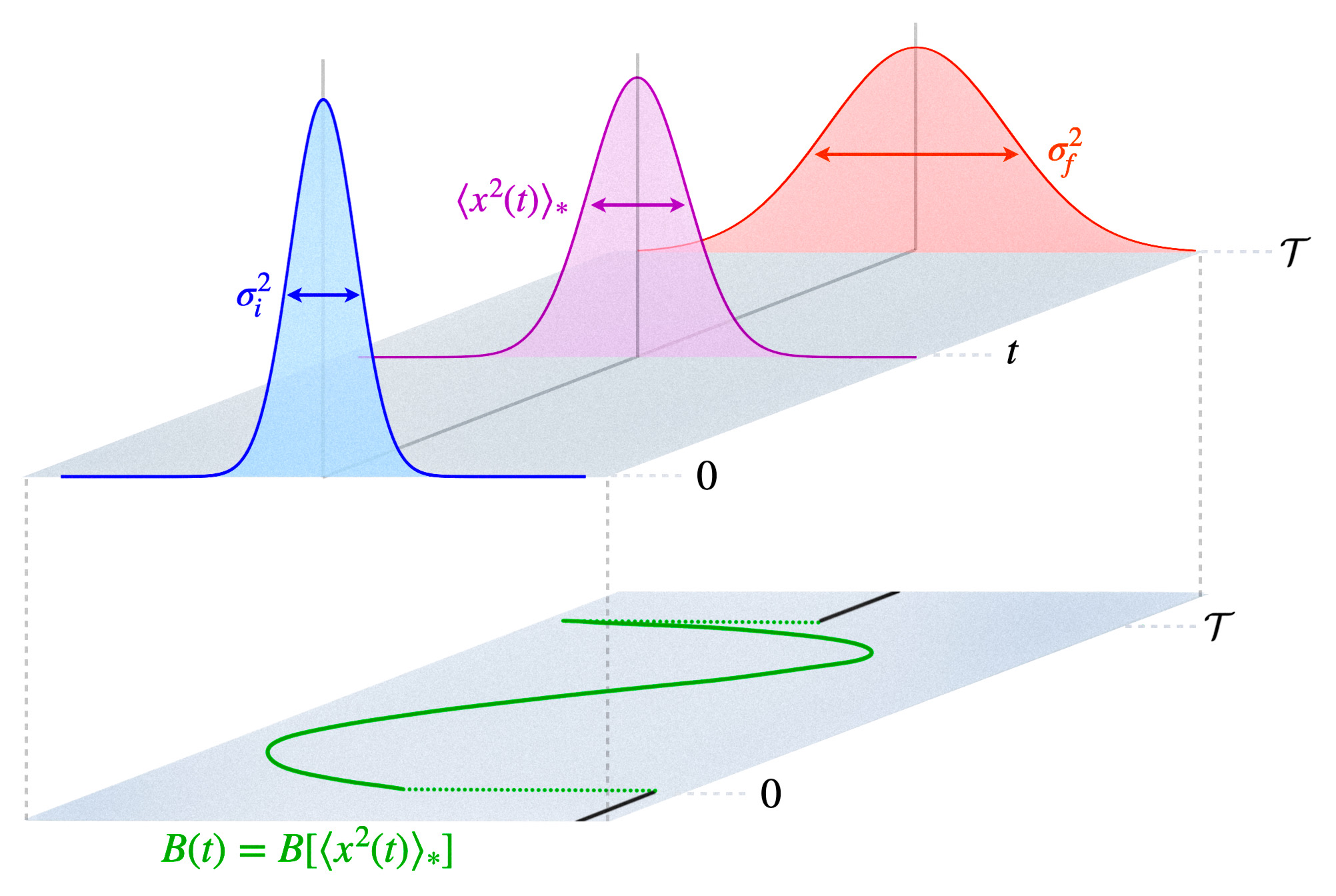}
    \caption{{Time-dependent modulation of the noise amplitude $B(t)$ of the self-propulsion forces is used to control non-equilibrium state-to-state transitions in time $\mathcal{T}$, connected by an inverse engineered mean squared displacement $\langle x^2(t) \rangle_*$. For example, transitions between a narrow "cold" state (blue) and a more dispersed "hot" state (red) can be studied.}}
    \label{fig:1}
\end{figure}

Recent work has highlighted a growing interest in controlling non‑equilibrium states and the transitions between them \cite{baldassarri2020engineered,prados2021optimizing,ilker2022shortcuts,zhong2022limited,engel2023optimal,patron2024minimum,gupta2025thermodynamic,monter2025optimal}. This naturally includes active systems, whose self-driven dynamics has made them a compelling setting to explore thermodynamic optimal control far from equilibrium \cite{olsen2025harnessingnonequilibriumforcesoptimize,gupta2023efficient,garcia2025optimal,schuttler2025active,davis2024active,soriani2025controlactivefieldtheories,casert2024learning,alvarado2026optimal,baldovin2026control,davis2026optimalmultiparametercontroltrapped,maurya2026optimaltransportactiveparticle,wang2025thermodynamic}.  As a complement to thermodynamic optimal control, inverse engineered approaches often referred to as shortcuts to adiabaticity, engineered swift equilibration or swift state-to-state transformations, provide another powerful tool to control state transitions \cite{le2016fast,guery2023driving,guery2019shortcuts,martinez2016engineered,chupeau2018engineered,lowen2026inverse}, which has recently been applied to active systems \cite{baldovin2023control,frim2023shortcut,cheng2026shortcutsstatetransitionsactive}. While many studies have examined control strategies that rely on externally imposed potentials, with or without combined modulation of activity, it is also useful to isolate the role of activity itself and consider control achieved solely through variations in self‑propulsion. { This limited regime of control in active systems isolates effects resulting purely from modulation of activity rather than combined control. Furthermore, this} viewpoint provides a non‑equilibrium counterpart of thermal control in passive systems, where rather than manipulating the physical temperature, one modulates the amplitude of the noise that drives self‑propulsion, thereby realizing a non‑equilibrium analogue of heating and cooling (see Fig. (\ref{fig:1})).

This level of control naturally arises in two distinct settings. First, across biological scales, modulation of self‑propulsion is ubiquitous, with organisms employing time‑dependent activity patterns ranging from intermittence \cite{friedrich2010high,polin2009chlamydomonas,gleiss2011convergent} to rapid bursts used for capture or evasion in predator–prey interactions  \cite{marras2015not,domenici2001scaling}. Second, recent experimental advances allow precise control of synthetic active particles, including activity tuning through optical methods \cite{buttinoni2012active,lozano2016phototaxis,rey2023light,4hrx-6rdq}.  Experimental
realizations in which a wide range of correlated noises can be digitally controlled offer platforms to test theories of activity-modulating systems\cite{goerlich2022harvesting}.

Here we propose a framework for studying activity-controlled swift state transitions in active systems, relating target states, activity control, and admissibility. After a general framework has been established, we consider transitions between stationary states. We further show that by allowing the initial state to be non-stationary, speed-ups beyond what is possible in passive systems can be reached, showing that the non-equilibrium nature of active systems may facilitate rapid effective cooling.

\section{Inverse engineered self-propulsion protocols}

Control strategies for active matter have recently been considered where self-propulsion or persistence time together with a trap stiffness are manipulated simultaneously \cite{baldovin2023control,davis2026optimalmultiparametercontroltrapped,frim2023shortcut}. Experimentally, { control over a
particle’s self-propulsion statistics alone} has also been realized \cite{goerlich2022harvesting}, {  offering a direct experimental arena to test the following framework}. Here we consider the level of control reached by solely tuning the active self-propulsion forces. As a model for our active particles, we consider active Ornstein-Uhlenbeck particles residing in a harmonic trap \cite{nguyen2022active}. Their stochastic dynamics, in the overdamped regime, takes the form
\begin{align}
    \dot x(t) &= \mu f_A(t) - \mu k x(t),\\
    \tau \dot f_A(t) &= - f_A(t) +\mu^{-1}\sqrt{2B(t)} \xi(t), \label{eq:fa}
\end{align}
where $k$ is the stiffness of the trap, $\mu$ the mobility, $\tau$ the persistence time of the active forces, and $B(t)$ is a time-dependent noise amplitude that determines the time dependence of  the active self-propulsion force $f_A(t)$.  We use the names self-propulsion protocol and noise protocol interchangeably when referring to $B(t)$. This choice of control naturally extends temperature protocols in passive systems, recovered in the $\tau\to 0$ limit. The noise $\xi(t)$ is a zero-mean delta-correlated Gaussian white noise.
We express the dynamics in terms of the velocity $p(t) =\dot x(t)$, which transforms the stochastic equations of motion into the form  \cite{bonilla2019active}
\begin{align}
    \dot x(t) &= p(t),\\
    \dot p(t) &= - \frac{p(t)+\mu k x(t)}{\tau}-\mu k p(t) + \sqrt{\frac{2B(t)}{\tau^2}}\xi(t).
\end{align}
The corresponding Fokker-Planck equation  for the joint probability density $\rho(x,p,t)$ reads
\begin{align}
    \partial_t\rho = -p\partial_x\rho +\partial_p \left[\left( \frac{1+\mu k \tau}{\tau} p +\frac{\mu k}{\tau} x\right) \rho + \frac{B(t)}{\tau^2} \partial_p\rho \right],
\end{align}
from which the dynamics of the particle follows.

\subsection{Inverse construction of a target mean squared displacement}
Traditionally, one specifies the dynamical rules and computes the mean squared displacement forward in time \cite{babel2014swimming,sharma2017brownian,jurado2024enhancement,gutierrez2025time}. Here, we consider the inverse situation: if an active particle, e.g., a living organism or a synthetic colloid, should reach a target mean squared displacement $\langle x^2(t) \rangle_*$, what time‑dependent self‑propulsion protocol $B(t)$ produces it?

From the above Fokker-Planck equation, we can derive the following closed set of equations for the second-order moments:
\begin{align}
    \partial_t \langle x^2\rangle &= 2\langle x p\rangle, \label{eq:x2}\\
    \partial_t \langle xp\rangle &= \langle p^2 \rangle -\frac{1+\mu k \tau}{\tau} \langle xp \rangle -\frac{\mu k}{\tau}\langle x^2 \rangle , \label{eq:xp}\\
    \partial_t\langle p^2\rangle &=-2\frac{1+\mu k\tau}{\tau} \langle p^2\rangle-\frac{2\mu k}{\tau} \langle xp \rangle +\frac{2B(t)}{\tau^2} .\label{eq:p2}
\end{align}
These relations can be used to derive an explicit differential relation between the noise protocol $B(t)$ and the second moment $\langle x^2 \rangle$. This can be obtained by taking three derivatives of the second spatial moment $\partial_t^3 \langle x^2 \rangle$ and iteratively use the above three equations until everything is expressed in terms of the mean squared displacement and the time-dependent noise strength. This results in a closed relation between the time dependent noise strength, and the desired mean squared displacement $\langle x^2(t) \rangle_*$, namely
\begin{equation}\label{eq:main}
B[\langle x^2\rangle_*]=   \frac{\tau^2}{4}\partial_t^3 \langle x^2 \rangle_* +a \partial_t^2 \langle x^2 \rangle_* + b \partial_t  \langle x^2 \rangle_* + c  \langle x^2 \rangle_* 
\end{equation}
where the (positive definite) coefficients are determined by the persistence time of the active forces and the trap strength:
\begin{align}
    a &= \frac{3\tau(1+\mu k \tau)}{4}\\
    b &= \tau\mu k +\frac{\tau^2}{2} \left(\frac{1+\mu k\tau}{\tau}\right)^2\\
    c &= \mu k(1+\mu k\tau)
\end{align}
Since the noise strength $B[\langle x^2\rangle_*]$ must be positive, we immediately have a constraint on admissible mean squared displacements which may be reached by this level of control, namely
\begin{equation}\label{eq:c11}
     \frac{\tau^2}{4}\partial_t^3 \langle x^2 \rangle_* +a \partial_t^2 \langle x^2 \rangle_* + b \partial_t  \langle x^2 \rangle_* + c  \langle x^2 \rangle_* \geq 0.
\end{equation}
As we will see later, this positivity constraint is closely related to speed limits when we seek protocols that connect non-equilibrium states. It is worth noting that Eq. (\ref{eq:main}) can be { recast} in the form of a classical mechanical problem where by $\langle x^2(t)\rangle_*$ is interpreted as a generalized coordinate and $B(t)$ a driving force. The problem then is formally equivalent to a driven harmonic oscillator with a third derivative ("jerk") in its equation of motion \cite{lowen2025gigantic}.

\subsection{Constraints on initial states}
To produce the desired mean squared displacement in time, the correct initial state must also be prepared. Indeed, if we write Eq. (\ref{eq:main}) as an operator relation 
\begin{equation}
    B[\langle x^2(t)\rangle_*]= \hat{\mathcal{L}}_t \langle x^2(t)\rangle_*,
\end{equation}
we can in principle solve this third-order equation for the mean squared displacement, which by construction takes the form
\begin{equation}
    \langle x^2(t)\rangle = \langle x^2(t)\rangle_* + c_1 f_1(t) + c_2 f_2(t) + c_3f_3(t),
\end{equation}
where $f_i(t)$ span the three-dimensional nullspace of the operator, i.e.,  $\hat{\mathcal{L}}_tf_j(t) = 0$. The three coefficients $c_j$ are determined by the initial state of the system, and when non-zero represent slowly relaxing transients on top of the desired dynamics. To exactly match the target mean squared displacement,  {  a necessary condition is the matching of derivatives $\partial_t^{0,1,2} \langle x^2(t)\rangle= \partial_t^{0,1,2} \langle x^2(t)\rangle_*$ at $t=0$}. Equivalently, from Eqs. (\ref{eq:x2})-(\ref{eq:xp}), these { derivative} constraints can be translated into the initial conditions on $\langle x^2(t) \rangle$, $\langle x(t) p(t) \rangle$ and $\langle p^2(t) \rangle$. For the zeroth derivative, clearly we must impose  $\langle x^2(0)\rangle=\langle x^2(0)\rangle_*$. For the first derivative matching, { $\partial_t \langle x^2(t)\rangle|_{t=0}= \partial_t \langle x^2(t)\rangle_*|_{t=0}$},  Eq. (\ref{eq:x2}) { gives for the initial correlations}
\begin{equation} \label{eq:c1}
    \langle x(0)p(0)\rangle = \frac{1}{2} \left [\partial_t \langle x^2(t) \rangle_*\right]_{t=0},
\end{equation}
and is uniquely determined by the initial slope of the target mean squared displacement. To reach a desired $\langle x^2(t) \rangle_*$ with a non-vanishing initial slope, the initial state must be prepared with correlations between position and velocity. Similarly, { $\partial_t^{2} \langle x^2(t)\rangle |_{t=0} = \partial_t^{2} \langle x^2(t)\rangle_*|_{t=0}$}, gives, through combining Eqs. (\ref{eq:x2})-(\ref{eq:xp}), the initial condition
\begin{align}\label{eq:c2}
  \langle p^2(0)\rangle  &=  \frac{1}{2}  \left [ \partial_t^2 \langle x^2\rangle_*\right]_{t=0} + \frac{1+\mu k \tau}{2\tau}  \left [  \partial_t \langle x^2\rangle_*\rangle\right]_{t=0} \nonumber \\
  & +\frac{\mu k}{\tau}\langle x^2(0) \rangle _*.
\end{align}
These conditions can be summarized in a covariance matrix for the phase space state $z=(x,p)$:
\begin{equation}
    \Sigma_0 \equiv \text{Cov}_0(z) = \begin{bmatrix}
\langle x^2(0) \rangle & \langle x(0) p(0)\rangle \\
\langle x(0) p(0)\rangle & \langle p^2(0) \rangle
\end{bmatrix}.
\end{equation}
This determines the initial state
\begin{equation}\label{eq:initialstate}
    \rho_0(z) = \frac{1}{2\pi \sqrt{\text{det}(\Sigma_0)}} \exp\left(-\frac{1}{2}z^T \Sigma_0^{-1} z\right).
\end{equation}
Since covariance matrices must be positive semi-definite $\text{det}(\Sigma_0)\geq 0$, this sets a constraint on the allowed mean squared displacement $\langle x^2(0)\rangle_*$, namely $\langle x(0) p(0)\rangle^2 \leq \langle x^2(0)\rangle_* \langle p^2(0)\rangle$. This is not simply a property of Gaussian states, but a general statistical fact; for any two variables $\langle xy\rangle^2$ is bounded from above  by $\sigma_x^2\sigma_y^2$ by H\"{o}lder's inequality \cite{durrett2019probability}. Explicitly, using the above relations, this constraint may be written in terms of the mean squared displacement as
\begin{align}
  & \frac{ \langle x^2(0) \rangle_*}{2}\left(   \left [ \partial_t^2 \langle x^2\rangle_*\right]_{0} + \frac{1+\mu k \tau}{\tau}  \left [  \partial_t \langle x^2\rangle_*\right]_{0} 
   \right) \geq\nonumber \\
   &\frac{1}{4} \left [\partial_t \langle x^2(t) \rangle_*\right]_{t=0}^2 -   \frac{\mu k}{\tau}\langle x^2(0) \rangle _*^2 .\label{eq:c22}
\end{align}
Equations (\ref{eq:c11}) and (\ref{eq:c22}) define the admissible space of mean squared displacements we can construct by solely controlling the particles activity.

\begin{figure}
    \centering
    \includegraphics[width=8.3cm]{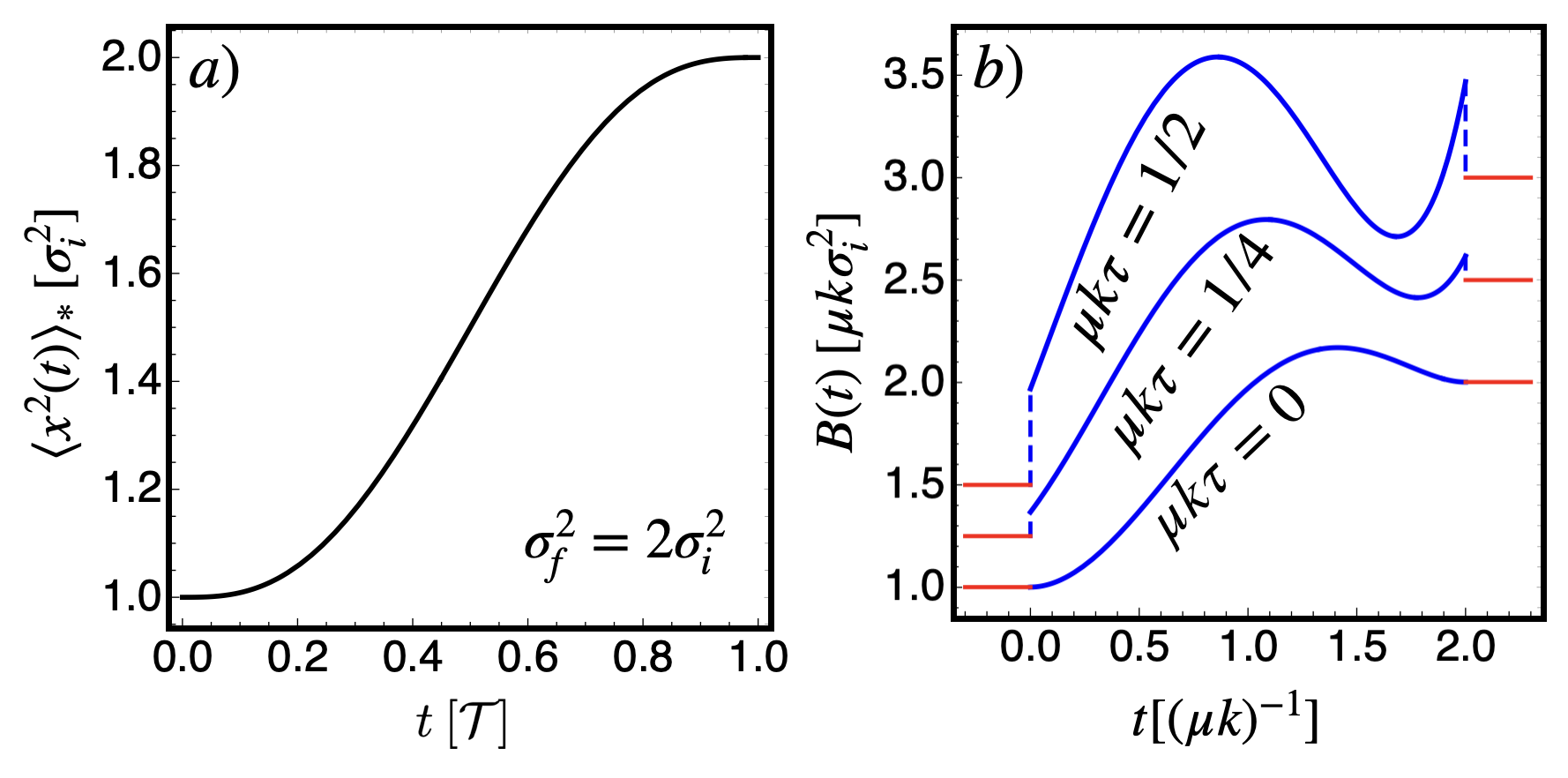}
    \caption{a) Imposed mean squared displacement, corresponding to the minimal polynomial ansatz in Eq. (\ref{eq:poly}), smoothly connecting two non-equilibrium steady states with { different} variance.  b) Activity protocols for various persistence { times}, ranging from the passive $\tau =0$ case where the noise simply performs a single overshoot, to more involved protocols for the active $\tau >0$ case. }
    \label{fig:msd}
\end{figure}

\begin{figure*}[t!]
    \centering
    \includegraphics[width=17.5cm]{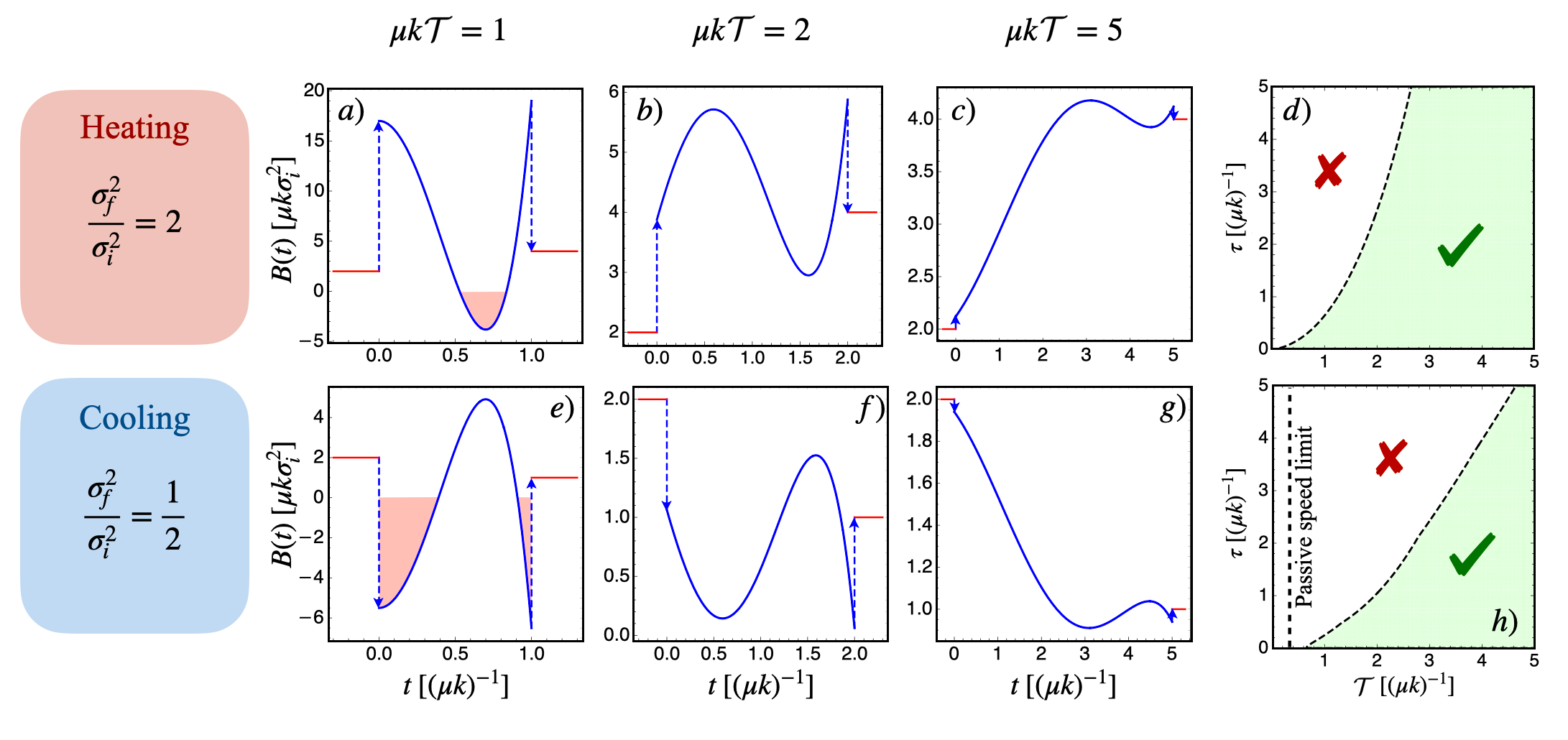}
    \caption{Noise protocols associated with a shortcut between two non-equilibrium steady states (red horizontal lines) in a harmonic trap. Protocols consist of discontinuities, non-monotonicity, and in some cases unphysical negative values. Positivity results in speed limits on the shortcuts, where the minimal shortcut time depends on the particles persistence time. Increasing variance (panels a-d) have the property that in the passive limit $\tau\to 0$ arbitrarily short protocol times are allowed. For decreasing variance (panels e-h) this is not the case, since the quench can not take place faster than the trap relaxation timescale. In panels a-c we set $\sigma_f^2/\sigma_i^2 = 2$ and in e-g we set $\sigma_f^2/\sigma_i^2 = 1/2$. Unless otherwise specified, all panels have $\mu k \tau= 1$. }
    \label{fig:bprotocols}
\end{figure*}

\section{Shortcuts between non-equilibrium steady states}
As a first application, we consider an active particle in a steady state with initial mean squared displacement $\sigma_i^2$. We aim to reach another steady state with mean squared displacement $\sigma_f^2$ in a transition time $\mathcal{T}$. In the passive limit $\tau \to 0$ the problem becomes one of temperature-controlled shortcuts \cite{pires2023optimal}. For temperature protocols with shortest possible transition time, so-called thermal brachistochrones, the shortest time can be identified through Pontryagin's principle, yielding bang-bang protocols of the temperature between its minimal and maximal values \cite{patron2022thermal}. For the case of one spatial dimension (or $d$-dimensional oscillators with isotropic potential), heating can be performed arbitrarily fast as long as there is no maximum temperature constraint. Cooling to a target variance $\sigma_f^2 < \sigma_i^2$ is fundamentally different since temperatures can not be negative, and must obey a speed limit \cite{prados2021optimizing, patron2022thermal}
\begin{equation}\label{eq:passive_limit}
    \mu k \mathcal{T} \geq \mu k \mathcal{T}_\text{min}^{(P)} = \log \frac{\sigma_i}{\sigma_f}
\end{equation}
This follows directly from the constraint that the temperature must remain positive, since the state variance obeys $  \sigma_f^2 = \sigma_i^2e^{-2\mu k \mathcal{T}} +2 \int_0^\mathcal{T} dt e^{-2\mu k (\mathcal{T}-t)} D(t) \geq  \sigma_i^2e^{-2\mu k \mathcal{T}}$. Saturating this corresponds to turning { off} any thermal fluctuations and re-activating them once the desired variance { is} reached, instantaneously equilibrating the system in its terminal state.

\begin{figure*}
    \centering
    \includegraphics[width=14cm]{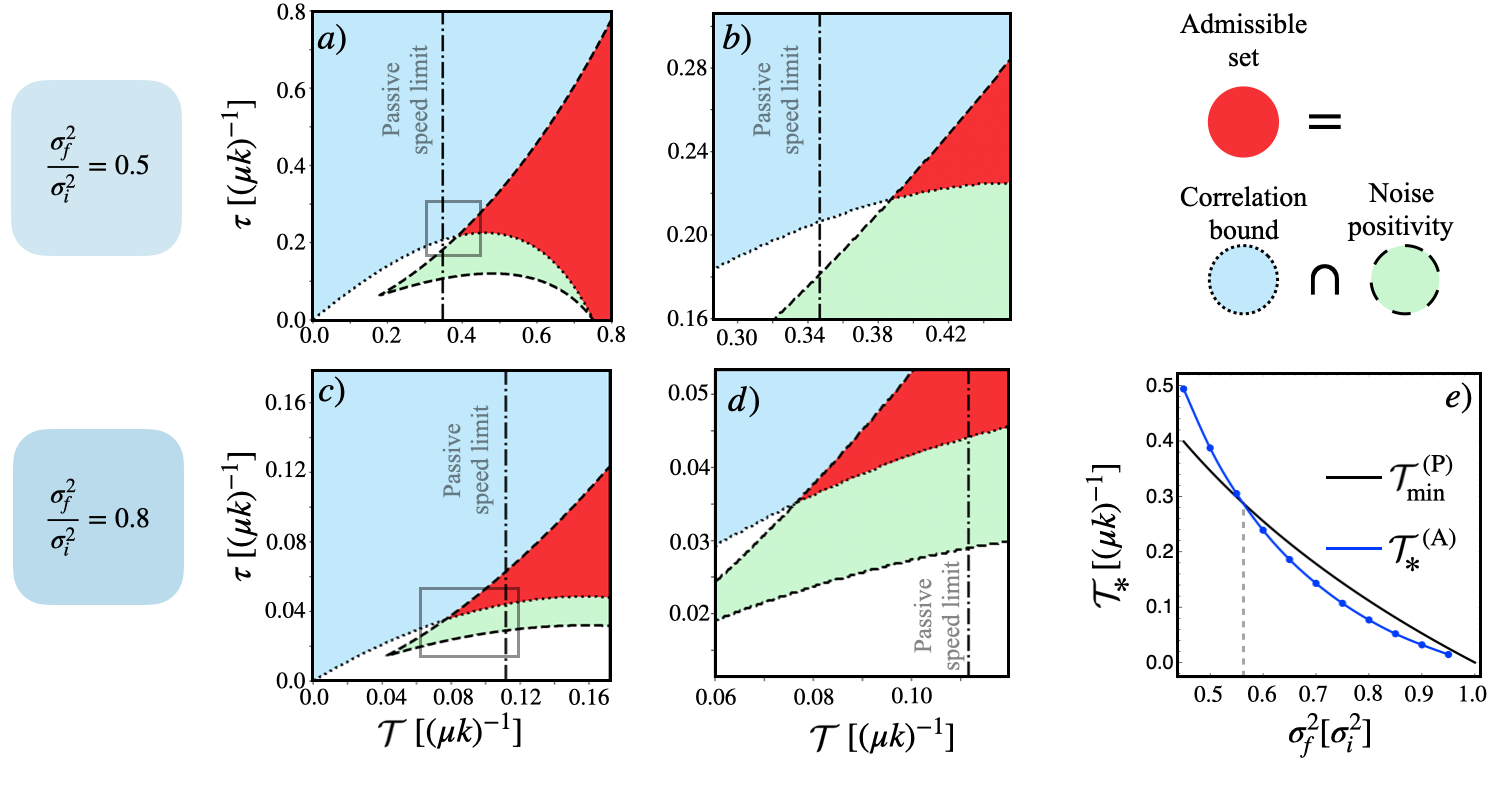}
    \caption{Admissible sets in parameter space $(\mathcal{T},\tau)$ where the correlation bound is satisfied (blue region with dotted boarders), and the noise positivity constraint is satisfied (green region with dashed boarders). Admissible set {(red regions)} corresponds to the intersection where both constraints are satisfied. Panels {a)} shows the case { $\sigma_f^2 = 0.5 \sigma_i^2$, with panel {b) zooming} in near the gray box in panel a). Dashed vertical line in { panels a) and b) shows} the passive speed limit. Panel c)-d) shows the case $\sigma_f^2 = 0.8 \sigma_i^2$, where we see that the admissible set reaches past the passive speed limit. Panel e) shows the minimal transition time allowed for the current protocol, obtained by picking the optimal persistence time that minimizes the transition time for each target variance. We see a transition from a region where active systems perform worse than a passive system, to a region where activity enhances cooling times.}}
    \label{fig:nonstationary}
\end{figure*}

In order for an active system to transition between two non-equilibrium steady states, { one} must also ensure that the internal degrees of freedom are made stationary. Concretely, for the active Ornstein-Uhlenbeck particle the constraints given by Eq. (\ref{eq:c1}) and Eq. (\ref{eq:c2}) must be satisfied, which in steady state take the values
\begin{align}
    \langle xp \rangle &= 0\\
    \langle p^2\rangle &= \frac{\mu k}{\tau} \langle x^2\rangle_*
\end{align}
These equations must be satisfied at times $t=0$ and $t=\mathcal{T}$. For these to be satisfied, Eq. (\ref{eq:c1}) and Eq. (\ref{eq:c2}) shows that we must require that the first two derivatives of the mean squared displacement vanish at the boundary of the protocol. Inspired from methods of engineered swift equilibration \cite{martinez2016engineered, pires2023optimal}, we make a polynomial { choice} for the mean squared displacement during the control protocol. A fifth-order polynomial of the form
\begin{equation}\label{eq:poly}
    \langle x^2(t) \rangle_* = \sigma_i^2 + \sum_{j=1}^5 c_j t^j
\end{equation}
is the lowest order polynomial which can be made to enforce the stationary constraints, yielding 
\begin{align}
    c_3 &= 10 \frac{\sigma_f^2-\sigma_i^2}{\mathcal{T}^3}\\
    c_4 &= - 15 \frac{\sigma_f^2-\sigma_i^2}{\mathcal{T}^4}\\
    c_5 &= 6 \frac{\sigma_f^2-\sigma_i^2}{\mathcal{T}^5}
\end{align}
with $c_1=c_2 = 0$. We note that for steady states, Eq. (\ref{eq:c22}) is automatically satisfied. Fig. (\ref{fig:msd}a) shows this mean squared displacement for a case where the non-equilibrium steady state variance is doubled, corresponding to effective heating, while Fig. (\ref{fig:msd}b) shows corresponding activity protocols obtained from Eq. (\ref{eq:main}). We see that while the passive noise protocol ($\tau = 0$) continuously connects to its corresponding equilibrium values, the active $(\tau >0)$ protocols { require} both discontinuous jumps and non-monotonicity to ensure that all degrees of freedom reach stationarity. Fig. (\ref{fig:bprotocols}a-c) shows the same protocols for various protocol durations $\mathcal{T}$, showing how short protocols are characterized by both larger discontinuities and stronger non-monotonic dependence on time, while slow protocols are less erratic. For very short protocols, the noise intensity can become negative during the control period, failing to satisfy Eq. (\ref{eq:c11}). {  The emergence of speed limits and non-monotonic protocols can be understood as a consequence of under-actuation \cite{zhong2022limited}: with only a single control variable, changes must propagate through the coupled dynamics before all degrees of freedom respond.} Fig. (\ref{fig:bprotocols}d) shows phase plots of admissible combinations of transition times $\mathcal{T}$ and persistence time $\tau${ , resulting from Eqs. (\ref{eq:c11}) and (\ref{eq:c22}),} showing that for every persistence time only sufficiently large transition times are allowed. This is naturally interpreted as a speed limit that constrains how fast a transition can take place between steady states out of equilibrium. Speed limits are well-documented for equilibrium to equilibrium shortcuts relying on combined manipulation of potential shape and temperature \cite{plata2020finite,ibanez2026cost} as well as for pure thermal control \cite{patron2022thermal,pires2023optimal}.  We see that in the equilibrium (passive) limit $\tau \to 0$, all $\mathcal{T}$ are allowed, signifying that in equilibrium systems the variance can be increased arbitrarily fast through noise control. 

Fig. (\ref{fig:bprotocols} e-g) shows complementary protocols where the control seeks to reduce the steady state variance,  corresponding to effective cooling. Yet again the protocols are discontinuous and non-monotonic. For these protocols however, the passive limit still has a speed limit, reflecting the fact that any decrease in variance can only rely on the inherent relaxation of the trap, obeying Eq. (\ref{eq:passive_limit}), shown in dashed vertical line in Fig. (\ref{fig:bprotocols} h).

\section{Rapid cooling through non-stationary initial states}
When considering transitions between non-equilibrium steady states, the requirement of stationarity at the protocol boundaries set strong constraints. Relaxing this requirement can drastically change the behavior of state to state transitions. Here we consider a system with initial variance $\sigma_i^2$, and consider protocols guaranteeing non-equilibrium steady state after a protocol duration $\mathcal{T}$ with final variance $\sigma_f^2\leq \sigma_i^2$. However, we do not require the initial state to be stationary, and study the consequences for cooling protocols.

The lowest-order polynomial that allows such transitions is given by 
\begin{equation}\label{eq:enhanced}
    \langle x^2(t) \rangle_* = \sigma_i^2 - 3\frac{\sigma_i^2-\sigma_f^2}{\mathcal{T}} t +3\frac{\sigma_i^2-\sigma_f^2}{\mathcal{T}^2} t^2 - \frac{\sigma_i^2-\sigma_f^2}{\mathcal{T}^3} t^3
\end{equation}
Here the initial state is clearly non-stationary, as seen from the presence of  non-zero initial correlations  $\langle x(0) p(0)\rangle = -\frac{3}{2}[\sigma_i^2-\sigma_f^2] /\mathcal{T}$. Positivity of the associated noise protocol, Eq. (\ref{eq:c11}), as well as the correlation bound in Eq. (\ref{eq:c22}), must be satisfied to ensure physical states.  Fig. (\ref{fig:nonstationary}) shows the admissible protocol durations and persistence times { for two different values of the final state variance. Fig. (\ref{fig:nonstationary}a), where { $\sigma_f^2 = 0.5 \sigma_i^2$}, shows the parameter space region where the correlation bound is satisfied (shaded light blue) and where the noise intensity is positive (shaded light green). The admissible set of parameters is given by their intersection (red shaded region). Fig. (\ref{fig:nonstationary}b) shows a zoom of the region indicated by the grey box in panel a), together with the passive speed limit represented by the vertical dashed line. {  We see that, in this case, the protocol does not cool the system faster than the passive speed limit.}
Fig. (\ref{fig:nonstationary}c-d) show the analogous plots for $\sigma_f^2 =0.8 \sigma_i^2$, in which case the admissible set { reaches beyond the passive speed limit.} Fig. (\ref{fig:nonstationary}e) shows the minimal transition time $\mathcal{T}_*^{(A)}$ for the active system, obtained by identifying the shortest possible protocol duration as a function of persistence time for each target variance $\sigma_f^2 \leq \sigma_i^2$. { We see a clear crossover from a region of target variances where enhanced cooling is not observed, to a regime where enhancement is observed. This indicates that the additional degrees of freedom associated with the non-equilibrium initial state can facilitate a rapid transition}. As a function of target variance, we see that { the crossover takes place} near { $\sigma_f^2 \approx 0.56 \sigma_i^2$}. While the particular behavior shown in Fig. (\ref{fig:nonstationary}) comes from the polynomial ansatz for the state variance, we expect the observed effects to persist to other protocols, {  and potentially be further enhanced if protocols are well chosen}. 

{ 
Our predictions can be tested using the experimental platform of Ref.\cite{goerlich2022harvesting}, where a colloidal particle is confined in a harmonic optical trap while a second laser generates a programmable stochastic force that mimics active fluctuations.  In particular, the speed-limit boundary in the $(\mathcal{T},\tau)$ plane and the { crossover in the optimal transformation time near $\sigma_f^2 \approx 0.56 \sigma_i^2$} (Fig. 4e) provide clear experimental signatures of the theory. Two experimental features are particularly important. First, short-duration protocols require sufficiently rapid modulation of the injected noise amplitude. Second, the setup must allow modulation over a sufficiently wide noise amplitude range, as the optimal protocols can require transient noise amplitudes that are up to an order of magnitude larger than the initial or final values.
}

A possible physical origin of the enhancement of cooling in active systems is the negative correlations between position and self-propulsion in the initial state that is needed to realize Eq. (\ref{eq:enhanced}), ensuring that particles are initially prepared in a way that is favorable to reduction in variance. These correlations can not be made arbitrarily negative however, as they are bounded by the fluctuations of the propulsion through $\langle x(0) p(0)\rangle^2 \leq \langle x^2(0)\rangle_* \langle p^2(0)\rangle$, i.e., Eq. (\ref{eq:c22}). Furthermore, negative correlations require initial negative slope in the mean squared displacement, which through Eq. (\ref{eq:c11}) may cause negative noise intensity. These bounds fundamentally { limit} the extent to which effective cooling can be performed in active systems.

{ In an experimental setting, initial states with negative position-velocity correlations can be prepared in several ways. Inspired by experimental approaches used to investigate the Mpemba effect, where artificial high-temperature Boltzmann states are generated through appropriate sampling schemes, one could realize the desired negative correlations by first preparing the system in its stationary state and then using a conditional sampling protocol: the particle orientation is measured, and the experiment is initiated only when the orientation points toward the trap center \cite{kumar2020exponentially}. Similar schemes have been implemented experimentally in the past in the context of Janus-type colloids \cite{bregulla2014stochastic}. {  While such procedures generally give rise to non-Gaussian states that do not take the form of Eq. (\ref{eq:initialstate}), the above framework remains valid for these initial states, as the moment hierarchy closes in the same way due solely to the structure of the equations of motion. Although such sampling protocols can give rise to a wide range of initial correlations, we expect the observed enhancement in cooling to qualitatively persist for sufficiently negative initial correlations.}
 }

Overall, these results suggest that engineering initial correlations may provide a viable strategy for designing faster cooling protocols in active matter systems. }

\section{Conclusions} 
Our results demonstrate that activity modulation alone can serve as a powerful control variable for active matter systems, serving as { an} active analogue of pure thermal control in passive systems. By identifying the admissible set of protocols shaped jointly by noise‑positivity constraints and correlation bounds, we clarify when and how target dynamical states can be reached. This framework enables the engineering of swift transitions between non‑equilibrium steady states through modulations of the statistics of self-propulsion alone. Allowing the initial state to be non‑stationary further expands the control landscape, revealing opportunities for accelerated transformations, including enhanced cooling that { outperforms} passive speed limits. {  This phenomenon may be conceptually related to the broader family of Mpemba-like anomalous relaxation effects, which demonstrate that relaxation dynamics can depend not only on macroscopic quantities such as the variance or effective temperature, but also on the non-equilibrium structure of the initial state \cite{kumar2020exponentially,bechhoefer2021fresh,teza2026speedups}. Our results therefore illustrate how non-equilibrium initial correlations can be harnessed to surpass passive speed limits, extending the general idea underlying Mpemba-like phenomena to controlled active systems.} Future works could consider optimal activity protocols that minimize transition time or thermodynamic costs, or extend the approach to other types of active motion.

\acknowledgments
KSO {acknowledges} support from the Alexander von Humboldt Foundation.


%

\end{document}